# Modeling and Real-Time Scheduling of DC Platform Supply Vessel for Fuel Efficient Operation

Kuntal Satpathi, *Student Member, IEEE*, VSK Murthy Balijepalli, *Member, IEEE*, and Abhisek Ukil, *Senior Member, IEEE*

*Abstract*—DC marine architecture integrated with variable speed diesel generators (DGs) has garnered the attention of the researchers primarily because of its ability to deliver fuel efficient operation. This paper aims in modeling and to autonomously perform real-time load scheduling of dc platform supply vessel (PSV) with an objective to minimize specific fuel oil consumption (SFOC) for better fuel efficiency. Focus has been on the modeling of various components and control routines, which are envisaged to be an integral part of dc PSVs. Integration with photovoltaic-based energy storage system (ESS) has been considered as an option to cater for the short time load transients. In this context, this paper proposes a real-time transient simulation scheme, which comprises of optimized generation scheduling of generators and ESS using dc optimal power flow algorithm. This framework considers real dynamics of dc PSV during various marine operations with possible contingency scenarios, such as outage of generation systems, abrupt load changes, and unavailability of ESS. The proposed modeling and control routines with real-time transient simulation scheme have been validated utilizing the real-time marine simulation platform. The results indicate that the coordinated treatment of renewable-based ESS with DGs operating with optimized speed yields better fuel savings. This has been observed in improved SFOC operating trajectory for critical marine missions. Furthermore, SFOC minimization at multiple suboptimal points with its treatment in the real-time marine system is also highlighted.

## Nomenclature

| | |
|---|---|
| $P_{mech}$ | Mechanical power at diesel engine (DE) shaft. |
| $P_{load}$ | Load demand. |
| $u_F$ | Fuel injection input signal. |
| $k_{pm}$ | Fuel injection system gain. |
| $\tau_{pm}$ | Fuel injection time constant. |
| $t_d$ | Dead-time of DE. |
| $J$ | DE rotor inertia moment. |
| $\omega_G$ | DE and generator angular speed. |
| $k_{loss}$ | DE rotational loss. |
| $T_G$ | Torque produced by the generator. |
| $p$ | Number of poles of generator. |
| $i_{Ms}, i_{Ts}$ | *M-T* axis current of wound rotor synchronous generator (WRSG) at stator flux reference frame (SFRF). |
| $\lambda_{Ms}, \lambda_{Ts}$ | *M-T* axis flux of WRSG at SFRF. |
| $T_T$ | Thrust developed by the thrusters. |
| $\tau_T$ | Torque developed by the thrusters. |
| $d_P$ | Propeller diameter. |
| $\omega_P$ | Speed of the propeller. |
| $P_T$ | Power developed by the thrusters. |
| $C_T$ | Thrust coefficient. |
| $C_\tau$ | Torque coefficient. |
| SOC | State of charge. |

Manuscript received March 5, 2017; revised June 2, 2017; accepted August 19, 2017. Date of publication August 24, 2017; date of current version September 15, 2017. This work was supported by the National Research Foundation Singapore through the Corporate Laboratory@University Scheme. (*Corresponding author: Kuntal Satpathi.*)

The authors are with the School of Electrical and Electronics Engineering, Nanyang Technological University, Singapore 639798 (e-mail: kuntal001@e.ntu.edu.sg; vsk@ntu.edu.sg; aukil@ntu.edu.sg).

Digital Object Identifier 10.1109/TTE.2017.2744180

## I. Introduction

PLATFORM supply vessel (PSV) plays a major role in the marine industry because of its ability to perform cruising and dynamic positioning (DP) operation [1]. Development of marine integrated power systems has enabled the marine loads esp. the propulsion systems to be powered from the common generation units [2], [3]. This resulted in the reduction of the number of installed prime movers and offered designers flexibility to place the generation system at any suitable location. The future operation of the marine vessels depends on the International Maritime Organization's air pollution requirements [4], [5]. To comply with the requirements, it is pertinent to develop fuel efficient marine vessels, hence limiting the exhaust gas emissions. In the conventional ac marine vessels, shaft electric machines have been proposed to minimize the fuel consumption [6]. DC marine systems are also proposed as they can operate with increased fuel efficiency as compared with the corresponding ac marine vessels [7], [8]. The lack of critical phase and frequency synchronizing parameters allows the interfaced DEs to run at variable speeds depending on total load demand, thus optimizing the specific fuel oil consumption (SFOC) and increasing the fuel efficiency [9]. Other advantages constitute the ease of integration with the energy storage systems (ESSs) [10]–[12], space and weight reductions [13], and lower losses as compared with the corresponding ac systems.

As analogous to the land-based multiterminal dc systems, dc marine vessels are envisaged to have a two-layer control system [14], [15]. The primary control system comprises of the dc bus voltage control [16] and active/reactive power control [17] with suitable protection and fault management algorithms [18]. The secondary control system comprises of the load forecasting, power flow algorithms, which will be executed for a given state and requirements of the marine missions [8], [19]. The secondary control system also helps in optimized power flow by curbing out unintended power consumption, hence minimizing the risk of blackout condition. It also aids in proper coordination of the load and generation system, which reduces the oversizing requirement of the generation systems.

The generation scheduling for commercial land-based power systems is usually done beforehand [8], which has been extensively studied for land-based microgrids [19] and electric vehicles [20]. Unlike land-based systems, the load profile of marine vessels is continuously changing owing to variable propulsion load demands. Thus, the generation output of marine vessels might be optimized for various operating scenarios. The optimized operation of the generation systems in the marine vessel is relatively new topic with a limited number of research attempts made in this area. Zahedi *et al.* [21] proposed the operation of DE at minimum SFOC with the help of integrated ESS where the authors have considered only the steady-state analysis to arrive a single point of minimized SFOC. Furthermore, much work on SFOC has been reported in the domain of ac marine vessels, such as shaft generator system integrated with the diesel generators (DGs) for possible minimization in fuel consumption [6], [22], agent-based real-time load management to control the loads [23], and stochastic approaches [15], [24]. One of the key research gaps in the past research studies has been the lack of modeling and implementation of realistic marine loads and marine missions. Although, the researchers have proposed various marine operating scenarios and SFOC minimization by considering mixed integer linear programming approach for unit commitment [25]; such approaches are NP-hard (nondeterministic polynomial) and may not be preferred choice for real-time scheduling system, which is the prime focus of this paper. Hence, this paper presents modeling and control of the various components of marine vessels incorporated in a real-time transient simulation scheme with dc optimal power flow (OPF) algorithms utilizing reduced bus-bar model for the scheduling of generation system. This paper considers SFOC minimization at multiple suboptimal points [26] as well as to incorporate the real-time dynamics of the dc shipboard systems. Furthermore, the performance of dc OPF-based secondary control algorithm and SFOC optimization with an option of ESS has been demonstrated for various marine missions.

This paper considers PSV as an example of marine vessel, which performs cruising operation for the logistics and DP operation to support the offshore supply vessels. Apart from the DGs, this paper also considers photovoltaic (PV)-based ESS [10], [27] as a part of generation systems of dc PSV. The focus is given on the integrated and automated generation operation by incorporating dc OPF-based algorithms in the proposed optimization framework to minimize the SFOC of DGs with the help of scheduling of ESS. The traditional analysis by offline simulations of the bigger and complex emerging dc marine systems is expected to consume longer time to generate results. This approach becomes quite cumbersome to analyze for the multiple test studies scenarios [28]. Thus, in this paper, the transient simulation framework for the entire dc shipboard system has been developed by utilizing the advantages of the real-time simulation platform. The study of the dynamics of the full shipboard power system along with the interaction of PV/ESS with the generation systems during contingencies has been carried out with the proposed approach. The real-time optimal power scheduling for generators and ESSs has been carried out for various marine missions under different contingencies, such as sudden load changes and network faults, to prove the efficacy of the proposed approaches. The trajectory of the SFOC of the DEs during the various contingencies has been studied, which could effectively be realized by the real-time simulation platform. Moreover, this method could be useful to validate the hardware design, converter control, and protection algorithms of the future dc marine vessels. Hence, the *smarter generation system* of the future dc PSV is proposed, which should be able to operate autonomously, encompassing both the primary and the secondary control system.

The rest of this paper is organized as follows. Section II covers the modeling of dc PSV, which comprises of the generation system and various marine loads. Section III comprises of the operating structure of the dc PSV considered in this paper. The transient simulation scheme with dc OPF algorithm handling ac/dc systems to minimize SFOC has been implemented in Section IV. The optimized generation scheduling applied with various marine operational scenarios is discussed in Section V. The real-time load scheduling is achieved using OPAL-RT OP5600-based simulator and this paper is concluded in Section VI.

## II. Modeling of DC Platform Supply Vessel

The bus-breaker model of the representative dc PSV is shown in Fig. 1. The model is in coherent with the commercially available PSVs [29], [30] with two-bus architecture to increase the reliability and survivability of the vessel. The representative vessel comprises of four DEs coupled with the synchronous generators ($\mathscr{P}^{\text{DG}}$) and one PV-based ESS ($\mathscr{P}^{\text{ESS}}$). The total generation capacity can be illustrated as per (1). The generators are interfaced with two level voltage source converter (2L-VSC) acting as an active front end rectifier and PV-based ESS is integrated via dc/dc converter for dc bus voltage and active/reactive power control

$$\mathbb{P}_{\text{Gen}} = \{\mathscr{P}^{\text{DG } n}, \mathscr{P}^{\text{ESS}} || n = 1:4\}. \tag{1}$$

The nominal bus voltage of dc PSV is set at 1500 $V_{\text{dc}}$ which according to the IEEE Std 1709-2010 falls under medium voltage dc shipboard architecture [31]. As compared with the land-based power systems, marine vessels have loads pertaining to different marine missions. Variable frequency propulsion system ($L_{\text{propulsion}}$) comprises of main propulsion (MP) systems to cater for the cruising loads ($\mathbb{L}^{CL}$); tunnel thrusters (TTs) and retractable thrusters (RTs) to cater for the DP load ($\mathbb{L}^{\text{DP}}$) [1]. The total connected propulsion load is illustrated as $L_{\text{propulsion}} = \{\mathbb{L}^{CL}, \mathbb{L}^{\text{DP}}\}$. The fixed frequency hotel loads are required for air conditioning/lighting systems, cranes/winches, small hotel motors, and so on. The hotel loads are classified into high-power ($\mathbb{L}^{\text{HL-high}}$) and low-power ($\mathbb{L}^{\text{HL-low}}$) loads. The miscellaneous loads ($\mathbb{L}^{\text{misc}}$) for radar and pulsed load operation are also considered as it may form the integral part of modern PSVs. The hotel loads can be illustrated as $L_{\text{hotel}} = \{\mathbb{L}^{\text{HL-high}}, \mathbb{L}^{\text{HL-low}}, \mathbb{L}^{\text{misc}}\}$ and the total load $L_{\text{total}}$ can be expressed as follows:

$$L_{\text{total}} = \{\mathbb{L}^{CL}, \mathbb{L}^{\text{DP}}, \mathbb{L}^{\text{HL-high}}, \mathbb{L}^{\text{HL-low}}, \mathbb{L}^{\text{misc}}\}. \quad (2)$$

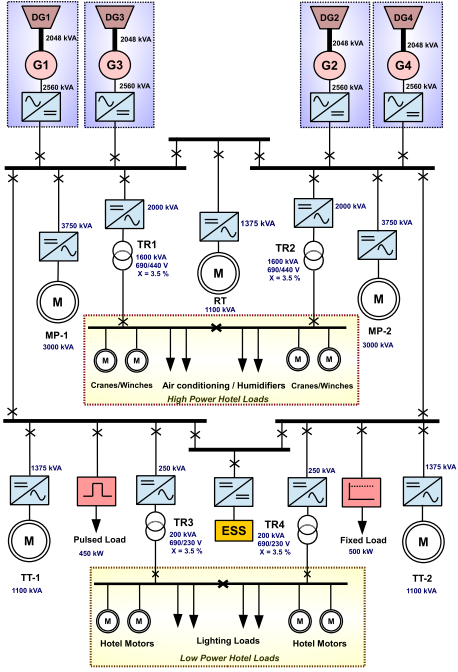

Fig. 1. Bus breaker model of representative dc PSV.

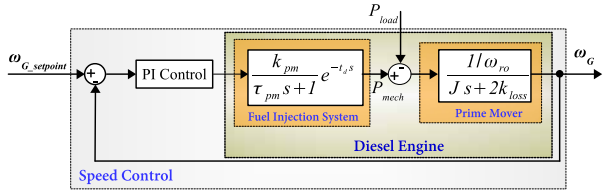

Fig. 2. Control loop diagram for the speed regulation of the DE.

The power rating of the components and converter systems is shown in Fig. 1. Derating factor of 125% has been used for the selection of the converters, cables, and bus bars to cater for the short time overload demands. The modeling and control of various components are discussed in Section II-A, B, C.

### A. Generation System Modeling

*1) Diesel Engine:* DEs are used as prime movers for the synchronous generators in dc PSV [3]. The prime-mover model comprises of fuel injection system, dead time ($t_d$) representing elapsed time until a cylinder produces torque and inertia of the rotating parts. The dead-time approximation of the prime-mover is realized by exponential delay and the transfer function is given in (3) [32], [33]. The differential equation governing the active power flow through the DE is

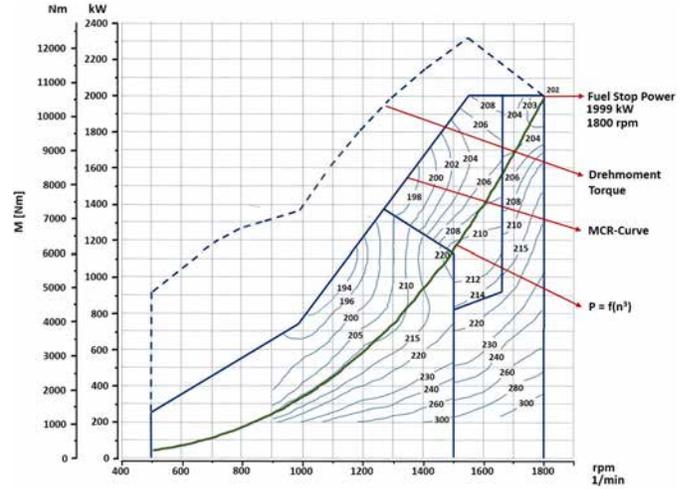

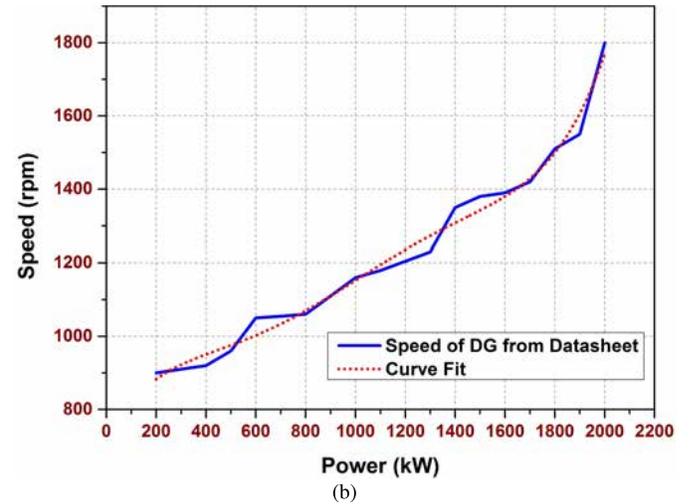

Fig. 3. (a) BSFC chart for representative 2000-kW DE [34] and (b) corresponding curve-fit of DE speed for optimized SFOC for various loading conditions.

shown in (4)

$$h_{\text{pm}}(s) = \frac{P_{\text{mech}}(s)}{u_F(s)} = \frac{k_{\text{pm}}}{\tau_{\text{pm}}s + 1} e^{-t_d s} \quad (3)$$

$$J \frac{d\omega_G}{dt} + k_{\text{loss}}\omega_G = \frac{P_{\text{mech}} - P_{\text{load}}}{\omega_G}. \quad (4)$$

Thus, the DE controls the synchronous generator by adjusting the mechanical power output. By linearization of the power flow equation (4) around the operating point $\omega_G = \omega_{Go}$, the transfer function reduces to

$$h_r(s) = \frac{\omega_G(s)}{\sum P} = \frac{1/\omega_{Go}}{Js + 2k_{\text{loss}}} \quad (5)$$

where $\sum P = P_{\text{mech}} - P_{\text{load}}$. The complete block diagram for DE speed control is shown in Fig. 2. In the dc PSV, DE should be able to operate at optimized SFOC by running at optimized speed. Fig. 3(a) represents the brake specific fuel consumption (BSFC) of the representative DE operating at different powers for different operating speeds [34]. The cost function of the optimized DE speed ($C(\omega)$) in terms of DG power output ($\mathscr{P}^{\text{DG}}$) is calculated to understand the operating points of the DE. This is achieved with the help of curve-fitting techniques and the cost function, $C(\omega)$ is derived as shown in the following:

$$C(\omega) = A_0 + A_1 \mathscr{P}^{\mathrm{DG}} + A_2 (\mathscr{P}^{\mathrm{DG}})^2 + A_3 (\mathscr{P}^{\mathrm{DG}})^3 + A_4 (\mathscr{P}^{\mathrm{DG}})^4 + A_5 (\mathscr{P}^{\mathrm{DG}})^5 \quad (6)$$

where $A_0 = 720.93$, $A_1 = 1.2591$, $A_2 = -0.00292$, $A_3 = 3.8104 \times 10^{-6}$, $A_4 = -2.1716 \times 10^{-9}$, and $A_5 = 4.5206 \times 10^{-13}$. The actual DE speed for various power requirements and the curve-fit version is shown in Fig. 3(b). The same cost function has been considered in the optimization framework presented in Section IV.

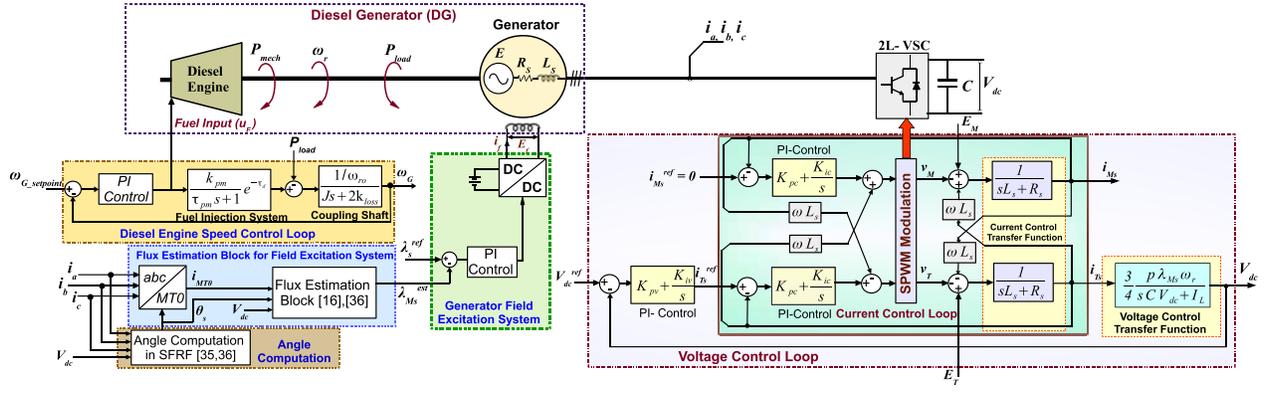

Fig. 4. Complete control loop of the DG interfaced with 2L-VSC.

*2) Diesel Generator–Rectifier Control:* DG includes full $M-T$ model of WRSG at stator flux reference frame (SFRF) with both the field and damper windings [35]. The output torque of the shipboard DG is maintained by independently controlling torque producing current, $i_{Ts}$ and machine flux, $\lambda_{Ms}$, which is achieved by maintaining $i_{Ms} = 0$ [16], [36]. With reference to [16], in this paper, the $M$-axis flux control is done by implementing flux estimation-based method in which the flux of the machine is estimated and suitably controlled for varying operating conditions [16], [36]. The dc bus voltage and line current control is realized by implementing PI regulator having bandwidth of 100 and 1000 Hz, respectively, such a range is suitable for simulating in the real-time operations as well. The switching frequency of the VSC is chosen to be 5 kHz and the VSC is modeled in both the average and detailed switching studies for comparative studies in the real-time simulation environment. The plant transfer function for dc bus voltage control is calculated by equalizing the input and output power flow while neglecting the line losses which is shown in (7). The plant transfer function for current control is calculated from the leakage inductance ($L_s$) and stator resistance ($R_s$) of the interfaced WRSG. The plant transfer functions for voltage control loop ($G_V$) and current control loop ($G_I$) are shown in (8). The combined control loop representation of the DG system interfaced with 2L-VSC is shown in Fig. 4

$$\mathscr{P}^{\mathrm{DG}} = T_G \cdot \omega_G \Longrightarrow 3p \frac{i_{Ts} \lambda_{Ms}}{4} \cdot \omega_r = C \frac{dv_{\mathrm{dc}}}{dt} \cdot v_{\mathrm{dc}} + v_{\mathrm{dc}} \cdot I_L \quad (7)$$

$$G_V = \frac{3}{4} \frac{p \lambda_{Ms} \omega_r}{s C V_{\mathrm{dc}} + I_L}, \quad G_I = \frac{1}{s L_s + R_s}. \quad (8)$$

## TABLE I
## LOAD PRIORITIES FOR DIFFERENT MARINE MISSIONS

| Marine Mission | Main Propulsion System | Tunnel Thrusters | Hotel Loads | Pulsed Loads |
|---|---|---|---|---|
| Cruising | HP | MP | MP | LP |
| Dynamic Positioning | HP | HP | MP | LP |
| Naval Warfare | HP | LP | LP | HP |
| At Port | HP | MP | MP | MP |

HP: High Priority, MP: Medium Priority, LP: Low Priority

### B. Marine Loads

DC PSV comprises of propulsion systems, thruster systems, hotel loads, and miscellaneous loads such as pulsed load to undertake different marine missions. Prioritization of the operation of these loads is dependent on the marine missions undertaken by the vessel [37] and a sample priority table considered in this paper is shown in Table I. Modeling of these loads is required to understand the power consumption pattern, which eventually be necessary for the scheduling of generation sources. The modeling and control of different loads are discussed in Section II-B2 and II-B3.

*1) Propulsion Systems:* In the PSVs, the propulsion systems ($L_{\mathrm{propulsion}}$) are the main consumers of energy, which undertake cruising and DP operation. The power requirement during the cruising operation ($\mathbb{L}^{CL}$) is dependent on the operating speed of the PSV ($\omega_P$) as shown in the following:

$$\mathbb{L}^{CL} \propto \omega_P^3. \quad (9)$$

The power requirement during DP operation ($\mathbb{L}^{\mathrm{DP}}$) carried out by PSVs is primarily dependent on the environmental forces and desired coordinate locations [38]. The sea current, wind velocity, surge, and sway of the vessels have to be balanced by the thrust produced by the thruster systems in order to maintain the desired coordinates. The generalized schematic of the DP system is shown in Fig. 5 [39]. The thrust production of the propeller is dependent on the speed ($\omega_P$), propeller geometry ($\alpha$), and hydrodynamic quantities ($\beta$). The thrust ($T$) and the torque ($\tau$) developed by the thrusters for speed ($\omega_P$) and diameter of the propeller ($d_P$) are given as follows [38], [39]:

$$T_T = g_T(n, \alpha, \beta) = C_T \rho d_P^4 \omega_P^2 \quad (10a)$$

$$\tau_T = g_\tau(n, \alpha, \beta) = C_\tau \rho d_P^5 \omega_P^2 \quad (10b)$$

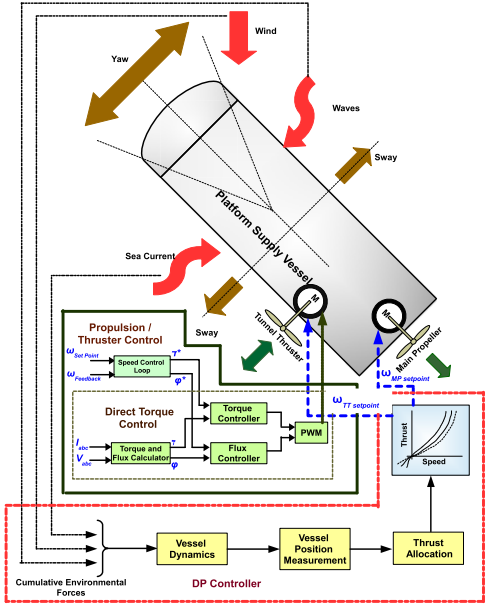

Fig. 5. Schematic of DP of PSV [38], [39].

| Parameters | Values |
|---|---|
| Irradiance | $1000\ W/m^2$ |
| Power | 305 W |
| Maximum Power Voltage ($V_{pm}$) | 54.7 V |
| Maximum Power Current ($I_{pm}$) | 5.58 A |
| Module Area | $1.63\ m^2$ |
| Total Available Area | $600\ m^2$ |
| Total Available PV Installation | $112\ kW \approx 110\ kW$ |
| Number of Modules | 360 |
| Modules in Series ($N_s$) | 6 |
| Modules in Parallel ($N_p$) | 60 |
| Rated Terminal Voltage ($V_{tm}$) | 330 V |

where $C_T$ and $C_\tau$ are determined by open-water tests for submerged vessels and are dependent on propeller advance velocity. In this paper, $C_\tau = 0.56$, density of water, $\rho = 997$ kg/m$^3$, and $d_P = 3.5$ m are considered. The power consumed by the thrusters of DP system is shown in the following:

$$\mathbb{L}^{DP} = 2\pi n\tau = C_\tau \rho d_P^5 \omega_P^3. \tag{11}$$

For the worst weather conditions, the power demanded by the DP system to maintain desired coordinates would be significantly higher than that of the power demand during the calm weather condition. Direct torque control is chosen over field-oriented control for propulsion motors and DP thrusters for its fast and superior performance and limited dependence on the machine parameters [40].

*2) Hotel Loads:* The percentage of the hotel loads ($L_{hotel}$) to the total load depends on the type of marine vessels. In PSVs, installed hotel load is much lower than the propulsion loads, but it is an important part of dc PSV. As shown in Fig. 1, two types of house loads are considered in this paper. The high power hotel loads ($\mathbb{L}^{HL-high}$) supplying power to cranes/winches and air-conditioning/humidifiers have a cumulative rating of 3200 kVA, 440 Vac, and operates at 60-Hz frequency. The low-power hotel loads ($\mathbb{L}^{HL-low}$) have a cumulative rating of 400 kVA, 230 Vac, and operates at 60-Hz frequency and are responsible for small hotel motors and lighting loads. Two level voltage source inverter with a constant output of 690 Vac, 60 Hz is utilized for the house loads.

*3) Miscellaneous Loads:* Miscellaneous loads ($\mathbb{L}^{misc}$) comprise of the pulsed and radar loads. Pulsed loads have presence in the modern naval vessels, which are used in the electromagnetic guns, free electron lasers, radars and high energy lasers, and draw huge amount of current lasting for short period of time. This intermittent nature of the loads has effect on the stability of the generation sources [41]. The pulsed load duration may vary from few microseconds to milliseconds.

In this paper, the pulsed load duration is selected to be 20 ms. The pulsed power load can be illustrated by the following:

$$\mathbb{L}^{pulse} = \frac{1}{T} \int_{t_1}^{t_2} P_o \cdot dt. \tag{12}$$

*C. Energy Storage System*

Future autonomous vessels are expected to be operated with different forms of renewables. Here, in this paper, it is assumed that battery-based ESS (BESS) units are supplied from the PV-based renewable source, which are interfaced with the dc PSV to cater for the short time load requirement and power fluctuations of the shipboard system. Such ESS also acts as reserve generation supply during the contingencies or sudden change of load. Unlike land-based power systems where energy is stored in the ESS when its cost is low and release the same when the grid electricity cost is high [19], here, there is no such variations in the cost. Electricity stored in the BESS is based on the marginal cost of the power supplied from the DGs. Cost function considered for the BESS is a constant price with the maximum power transfer based on the state of SOC as depicted in the following:

$$C_{ess} = f_p / kwh. \tag{13}$$

The operational capacity of the ESS is restricted to 10% of the total installed generation capacity as illustrated in the following:

$$\mathscr{P}^{ESS} = 0.1 * \mathbb{P}_{Gen}. \tag{14}$$

The PV-BESS is set to operate in optimal mode by limiting the battery usage between 20% and 100% of total storage capacity; furthermore, the scheduling constraints are imposed accordingly. The selection, sizing, and schematic of the PV-BESS are described in the following.

*1) PV Energy Sources in Marine Vessels:* As per the green ship initiative, combination of PV-DG-based generation systems is expected to be part of future marine vessels [10]. The rating and capacity of the PV panel are dependent on the available space in the target marine vessel [27]. For marine vessels undertaking longer voyages, e.g., liquefied natural gas carriers having easier accessibility to roof top terrace; proliferation of PV-based generation system with 20% capacity of the total generation system has been suggested [10], [27], [42]. This paper considers the PSV with size and power rating comparable with the commercial available PSVs, such as Rolls-Royce UT776 [29] and Viking Queen [30]. These two

| | Parameters | Values |
|---|---|---|
| Battery Module | Model Name | SAFT Seanergy Modules |
| | Nominal Voltage | 46.2 V |
| | Nominal Capacity | 60 Ah |
| Battery Pack | Nominal Voltage | 650 V |
| | Nominal Capacity | 1200 Ah |
| | Modules in Series ($N_s$) | 14 |
| | Modules in Parallel ($N_p$) | 20 |

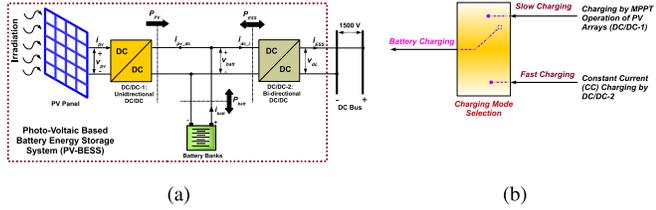

Fig. 6. (a) Schematic of PV-based BESS (PV-BESS) and (b) representation of the battery charging schemes.

PSVs are equal in size and rating according to the description provided in the whitepaper [29], [30]. The deck area of the PSV is of commercial interest and cannot be utilized for PV installation [27]. Thus, the PSV has limited available space for PV array installation. It has been assumed that the 600 m² of the total available area of 1800 m² [29], [30] is available for installing PV arrays. Considering the parameters of the commercially available Sunpower 305 Solar Panel [43] with the available installation area of 600 m², the rating of total installed PV array is described in Table II. The economic analysis of the PV panel is dependent on several market parameters and specifications, but it is expected to be consistent with the method provided in [42]. The comprehensive cost analysis of PV panels in dc PSV is currently beyond the scope of this paper.

*2) Sizing of PV-BESS System:* According to the IEEE Std 1562-2007 [44] and the IEEE Std 1013-2007 [45], the sizing of the BESS connected to the PV is determined while assuming that there is no power available from the PV system. The BESS is installed in the dc Shipboard Power System with the intention of fulfilling the intermittent loads and support the generation system during various contingencies [8]. The PV power is primarily used to charge the BESS and maintain its SOC at maximum possible level. The selection of the BESS has been done to minimize the weight and size constraints of the dc marine vessels [8]. Furthermore, the 10% power level of BESS is chosen to make it consistent with the trends of BESS selection in commercially available marine vessels [46]. The parameters of the BESS is chosen according to the commercially available *SAFT Seanergy* modules, which are suitable for hybrid propulsion applications [47]. The parameters of the battery module and the battery pack considering 10% of power demand are shown in Table III.

*3) Schematic of PV Interfaced BESS:* The schematic of the PV and BESS interfaced with the dc bus is shown as per Fig. 6(a) [48]. To extract the maximum power, the PV panel is interfaced with the unidirectional dc/dc-1 converter which works on perturb and observe ($P\&O$)-based maximum power point tracking (MPPT) algorithm [49] for the proposed real-time transient simulation scheme. The modeling of the PV generator and the MPPT algorithm is consistent with strategy discussed in [49]. The dc/dc-2 converter is a bidirectional converter used to interface PV-BESS system to the dc bus. The modeling and control of dc/dc-2 is consistent with the approach provided in [7]. During the normal operation, when the SOC of the battery is above threshold limit ($SOC_{max}$), the dc/dc-2 operates at boost conversion mode supplying the power to the dc ship as fulfilling scheduled generation requirements and complying with (15a)–(15h). The variables in (15a)–(15h) are consistent with annotations shown in Fig. 6(a)

$$P_{PV} > 0 \tag{15a}$$
$$P_{Batt} > 0 \tag{15b}$$
$$P_{ESS} > 0 \tag{15c}$$
$$i_{PV\_dc} > 0 \tag{15d}$$
$$i_{batt} > 0 \tag{15e}$$
$$i_{dc\_i} = i_{batt} + i_{PV\_dc} > 0 \tag{15f}$$
$$i_{ESS} > 0 \tag{15g}$$
$$P_{ESS} = P_{batt} + P_{PV}. \tag{15h}$$

When the SOC of the BESS is below lower threshold ($SOC_{min}$), it could either be charged exclusively by the PV system or by the combination of PV system and dc/dc-2 converter. Since the power output of the PV array has limitation owing to dependence on available irradiation, charging with PV panel would result in slow charging as shown in Fig. 6(b). Constant current-based fast charging of the BESS can be carried out by maintaining output current of dc/dc-2 at desired charging rate suggested by the manufacturers. During the charging operation supported by both PV panels and dc/dc-2, (16a)–(16h) are satisfied

$$P_{PV} > 0 \tag{16a}$$
$$P_{Batt} < 0 \tag{16b}$$
$$P_{ESS} < 0 \tag{16c}$$
$$i_{PV\_dc} > 0 \tag{16d}$$
$$i_{batt} < 0 \tag{16e}$$
$$i_{dc\_i} = i_{batt} - i_{PV\_dc} < 0 \tag{16f}$$
$$i_{ESS} < 0 \tag{16g}$$
$$P_{batt} = P_{ESS} + P_{PV}. \tag{16h}$$

## III. OPERATION OF DC PLATFORM SUPPLY VESSEL

Installed generation capacity of the PSV is generally lower than the total loads connected to the system. This is because a defined set of loads are activated for particular marine mission as indicated in Table I. For cruising operation, PSV operates mostly in fixed speed condition and for DP mode, the operating speed may change depending on the environmental conditions. Hence, the brake power of PSV for DP operation is not constant as shown in Fig. 7, where the vessel mostly operates between 20% and 80% of the total installed brake power [34]. Thus, incorporating dc OPF algorithm and operating the vessel at minimum SFOC can be implemented to increase the fuel efficiency. For the dc OPF reduced bus-branch model, segregating the ac and dc subsystems with ac/dc-dc/ac boundary node is shown in Fig. 8. Generators Gen-1 and Gen-3 are clubbed together and are connected to bus B1. Similarly, the Gen-2 and Gen-4 are clubbed together and connected to bus B2. The total generation capacity ($\mathbb{P}_{Gen}$) of the DGs and the ESS with the bus they are interfaced to is illustrated in the following:

$$\mathbb{P}_{Gen} = \{\mathscr{P}_{B1}^{Gen13}, \mathscr{P}_{B2}^{Gen24}, \mathscr{P}_{B14}^{ESS}\}. \quad (17)$$

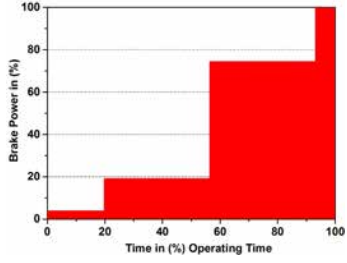

Fig. 7. Brake power of PSV for DP operation.

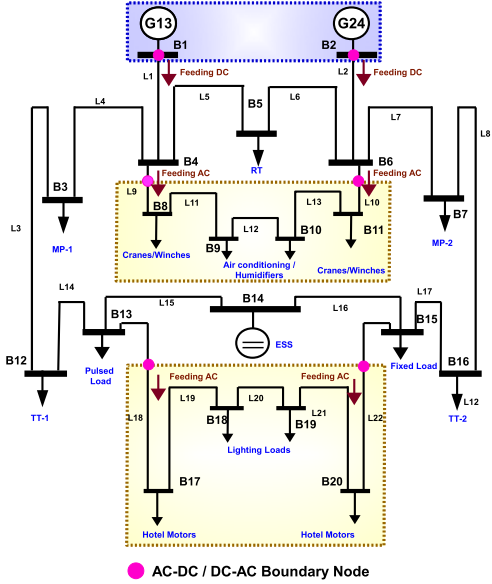

Fig. 8. Reduced bus-branch model of representative dc PSV.

| To Bus | From Bus | Line | R (mΩ) | X (mΩ) | P (kVA) | V (pu) |
|---|---|---|---|---|---|---|
| 4 | 1 | L1 | 0.48 | -na- | 2x2048 | 1.0 |
| 6 | 2 | L2 | 0.48 | -na- | 2x2048 | 1.0 |
| 12 | 3 | L3 | 0.092 | -na- | 2x1750 | 1.0 |
| 4 | 3 | L4 | 0.092 | -na- | 2x1750 | 1.0 |
| 5 | 4 | L5 | 1.5 | -na- | 3000 | 1.0 |
| 6 | 5 | L6 | 1.2 | -na- | 3000 | 1.0 |
| 7 | 6 | L7 | 0.64 | 0.75 | 1600 | 1.0 |
| 16 | 7 | L8 | 0.64 | 0.75 | 1600 | 1.0 |
| 9 | 4 | L9 | 3.2 | -na- | 1100 | 1.0 |
| 11 | 6 | L10 | 3.2 | -na- | 1100 | 1.0 |
| 9 | 8 | L11 | 2.56 | -na- | 450 | 1.0 |
| 10 | 9 | L12 | 2.56 | -na- | 450 | 1.0 |
| 11 | 10 | L13 | 2.56 | 2.31 | 2x200 | 1.0 |
| 13 | 12 | L14 | 2.56 | 2.31 | 2x200 | 1.0 |
| 14 | 13 | L15 | 3.2 | -na- | 1100 | 1.0 |
| 15 | 14 | L16 | 0.03 | -na- | 2x2048 | 1.0 |
| 16 | 15 | L17 | 0.03 | -na- | 2x2048 | 1.0 |
| 17 | 13 | L18 | 0.03 | -na- | 2x2048 | 1.0 |
| 18 | 17 | L19 | 0.03 | -na- | 2x2048 | 1.0 |
| 19 | 18 | L20 | 0.03 | -na- | 2x2048 | 1.0 |
| 20 | 19 | L21 | 0.03 | -na- | 2x2048 | 1.0 |

| Bus | $P_{max}\uparrow$(kW) | $P_{min}\downarrow$(kW) | Q(kVAr) | V(pu) |
|---|---|---|---|---|
| B1 | +4096 | 0 | -na- | 1.05 |
| B2 | +4096 | 0 | -na- | 1.05 |
| B3 | -3000 | 0 | -na- | 1 |
| B4 | -640 | 0 | -480 | 1 |
| B5 | -1100 | 0 | -na- | 1 |
| B6 | -640 | 0 | -480 | 1 |
| B7 | -3000 | 0 | -na- | 1 |
| B8 | -240 | 0 | -180 | 1 |
| B9 | -400 | 0 | -300 | 1 |
| B10 | -400 | 0 | -300 | 1 |
| B11 | -240 | 0 | -180 | 1 |
| B12 | -1100 | 0 | -na- | 0.95 |
| B13 | -450 | 0 | -na- | 0.95 |
| B14 | +820 | 0 | -na- | 1 |
| B15 | -450 | 0 | -na- | 0.95 |
| B16 | -1100 | 0 | -na- | 0.95 |
| B17 | -80 | 0 | -60 | 1 |
| B18 | -80 | 0 | -60 | 1 |
| B19 | -80 | 0 | -60 | 1 |
| B20 | -80 | 0 | -60 | 1 |

The loads $\mathbb{L}_{CL}$, $\mathbb{L}_{DP}$, $\mathbb{L}_{HL_{high}}$, $\mathbb{L}_{HL_{high}}$, and $\mathbb{L}_{misc}$ interfaced with respective bus are depicted in the following:

$$\mathbb{L}_{CL} = \{\mathscr{L}_{B3}^{MP1}, \mathscr{L}_{B7}^{MP2}\} \quad (18a)$$

$$\mathbb{L}_{DP} = \{\mathscr{L}_{B12}^{TT1}, \mathscr{L}_{B16}^{TT2}, \mathscr{L}_{B5}^{RT}\} \quad (18b)$$

$$\mathbb{L}_{HL_{high}} = \{\mathscr{L}_{n}^{HL1}||n = B8 \text{ to } B11\} \quad (18c)$$

$$\mathbb{L}_{HL_{low}} = \{\mathscr{L}_{m}^{HL2}|| \ m = B17 \text{ to } B20\} \quad (18d)$$

$$\mathbb{L}_{misc} = \{\mathscr{L}_{B13}^{PL}, \mathscr{L}_{B15}^{FL}\}. \quad (18e)$$

## IV. PROPOSED REAL-TIME DC PSV POWER MANAGEMENT SYSTEM

### A. Problem Statement

As explained in Section I, operation of dc PSV demands real-time scheduling mechanisms to tackle different loads and generation sources in various operating conditions. The real-time transient simulation scheme with dc OPF is more suitable for such conditions, which determines the power injections of the DGs and ESS to minimize the SFOC in real time, subjected to physical and operational constraints (relevant data in Tables IV and V). Equality constraints include power balance at each node and inequality constraints include the network operating limits, DG limits, ESS limits, and limits on the other control variables. These control variables include active power output of the generators, power electronic controls, amount of load disconnected, and the status of storage devices. Hence, subsequent to the modeling of dc PSV, real-time dc OPF with the objective of minimizing SFOC considering all control and state variables with real-time optimization framework is the objective behind this paper.

### B. Problem Formulation

The real-time transient simulation system for the generation scheduling of the dc PSV is governed in such a way to effectively utilize the available resources onboard to minimize the SFOC of the DGs. In this process, the optimization considers the set constraints as follows:

$$\text{Minimize SFOC } F(\{\mathscr{P}_{B1}^{\text{Gen13}}, \mathscr{P}_{B2}^{\text{Gen24}}, \mathscr{P}_{B14}^{\text{ESS}}\}) = f(x, u) \quad (19)$$

$$\text{s.t. } w(x, u) = 0 \quad (20)$$

$$q(x, u) \leq 0 \quad (21)$$

where the cost function referring to the active power of the energy resources is minimized while respecting the equality constraints $w(x, u)$ and inequality constraints $q(x, u)$. These constraints can be viewed as linear and nonlinear constraints

$$w(x, u) = \begin{bmatrix} w_{\text{nl}}(x, u) \\ J_e(x, u) + o_e \end{bmatrix} \quad (22)$$

$$q(x, u) = \begin{bmatrix} q_{\text{nl}}(x, u) \\ J_i(x, u) + o_i \end{bmatrix} \quad (23)$$

where $J_e$ and $J_i$ are constants and need to be calculated only once. State variables of converter and dc network are set up to this framework. Energy storage and dc side converter power feed-in are mapped to the corresponding ac buses to satisfy the Kirchhoff law.

The vector $x$ consists of dependent variables such as fixed parameters such as reference angles, noncontrolled generator or ESS outputs, noncontrolled loads, and line parameters. The vector $u$ consists of control variables, including real power generation, PSV load shedding parameters/priorities, ESS charging and discharging limits, ramp rates of the DG, dc line flows, and converter control settings. The equality and inequality constraints are, namely, power flow equations, limits on all control variables, generation/load balance, branch flow limits, and SOC limits. Considering Fig. 8, which represents the reduced bus-bar model of dc PSV (having DGs, ESS, and different types of loads), for an anticipated group of loads, total system generation should be scheduled in such a way to minimize the SFOC of DG. In such cases, the network equality constraints are represented by the standard load flow equations [14]. PSV load balance equation is as follows in the real-time operation:

$$\sum_{i=1}^{B} (\mathscr{P}_{\text{B}i}^{\text{Gen}} + \mathscr{P}_{\text{B}i}^{\text{ESS}}) - \sum_{i=1}^{D} (\mathscr{P}_{\text{D}i}^{L} + \mathscr{P}_{\text{D}i}^{\text{ESS}}) - \mathscr{P}_{\text{Losses}} = 0. \quad (24)$$

Inequality constraints limits are set accordingly, for example, generator limits are set as

$$\mathscr{P}_{\text{B}i_{\text{min}}}^{\text{Gen}} \leq \mathscr{P}_{\text{B}i}^{\text{Gen}} \leq \mathscr{P}_{\text{B}i_{\text{max}}}^{\text{Gen}} \quad (25)$$

$$Q_{\text{B}i_{\text{min}}}^{\text{Gen}} \leq Q_{\text{B}i}^{\text{Gen}} \leq Q_{\text{B}i_{\text{max}}}^{\text{Gen}}. \quad (26)$$

Load shedding or load balancing limits have been set as

$$0 \leq \mathscr{L}_{\text{B}i}^{\text{shed}} \leq \mathscr{L}_{\text{B}i}^{D_{\text{total}}}. \quad (27)$$

Energy storage limits set as

$$\mathscr{P}_{\text{B}i_{\text{min}}}^{\text{ESS}} \leq \mathscr{P}_{\text{B}i}^{\text{ESS}} \leq \mathscr{P}_{\text{B}i_{\text{max}}}^{\text{ESS}}. \quad (28)$$

Converter voltage limits on the ac side are nonlinear in nature and can be set as

$$\mathscr{V}_{\text{conv}_{\text{min}}}^{2} \leq \mathscr{V}_{\mathscr{R}_{\text{conv}}}^{2} + \mathscr{V}_{\mathscr{I}_{\text{conv}}}^{2} \leq \mathscr{V}_{\text{conv}_{\text{max}}}^{2}. \quad (29)$$

Converter filter side constraints are as follows, where $V_R$ and $V_I$ are real and imaginary parts of the voltage:

$$\mathscr{V}_{\text{filter}_{\text{min}}}^{2} \leq V_{R_{\text{filter}}}^{2} + V_{I_{\text{filter}}}^{2} \leq \mathscr{V}_{\text{filter}_{\text{max}}}^{2} \quad (30)$$

and the limits on the converter current and dc voltages are linear in nature and are as follows:

$$\mathscr{V}_{\text{min}}^{\text{dc}} \leq \mathscr{V}^{\text{dc}} \leq \mathscr{V}_{\text{max}}^{\text{dc}} \quad (31)$$

$$I_{\text{min}}^{\text{conv}} \leq I^{\text{conv}} \leq I_{\text{max}}^{\text{conv}}. \quad (32)$$

$q(x, u)$ in (23) is formed by (25)–(32) as stated previously. It is to be noted that the voltage and branch flow limits are the only nonlinear limits on the ac side. Options of setting branch limits, and other operational limits in the dc PSV are implemented as well, but skipped for better readability of this paper.

Power generation schedules obtained from the optimization framework have been fed to the SFOC calculation block to calculate the speed ($C(\omega)$) based on the equation derived in Section II, where speed is derived as a function of generator schedules

$$C(\omega) = f(\mathscr{P}^{\text{DG}}). \quad (33)$$

SFOC at each optimized speed point corresponding to the power schedules has been calculated using the *SFOC lookup table* available in the SFOC calculation block as shown in Fig. 9(a).

## C. Real-Time Transient Simulation of DC PSV

The architecture of the real-time transient simulation setup comprising of generator scheduling scheme based on dc OPF is shown in Fig. 9(a). Depending on the operating mode, the scheduling block takes the input from the operating personnel. The available generation is also fed to calculate the reserve generation capacity and setting the upper limits of the generation system. Load estimation is the critical step where the rate of the load changes has been assessed and passed as an input to the proposed algorithm. Tabulation methods for electrical loads during marine missions which are proposed in [50] are adopted in this paper. Line parameters (Table IV) and bus-bar parameters (Table V) are fed to ensure that the loading in each line/bus stays within prescribed limits. The scheduled generation is calculated and fed to the SFOC optimization block to calculate the power demand and corresponding speed set point of each of the in-line DGs. The scheduling block and the SFOC algorithm are the part of the controller, while the rest of the system of dc PSV system is divided into master/slave computational subsystems and loaded into computing cores of OPAL-RT OP5600-based real-time simulator. The segregation of cores of the real-time simulation model has been realized with the help of gyrators and the partitioning contours of the divided subsystems are shown in Fig. 9(a). The description of the real-time simulation system is described in the following.

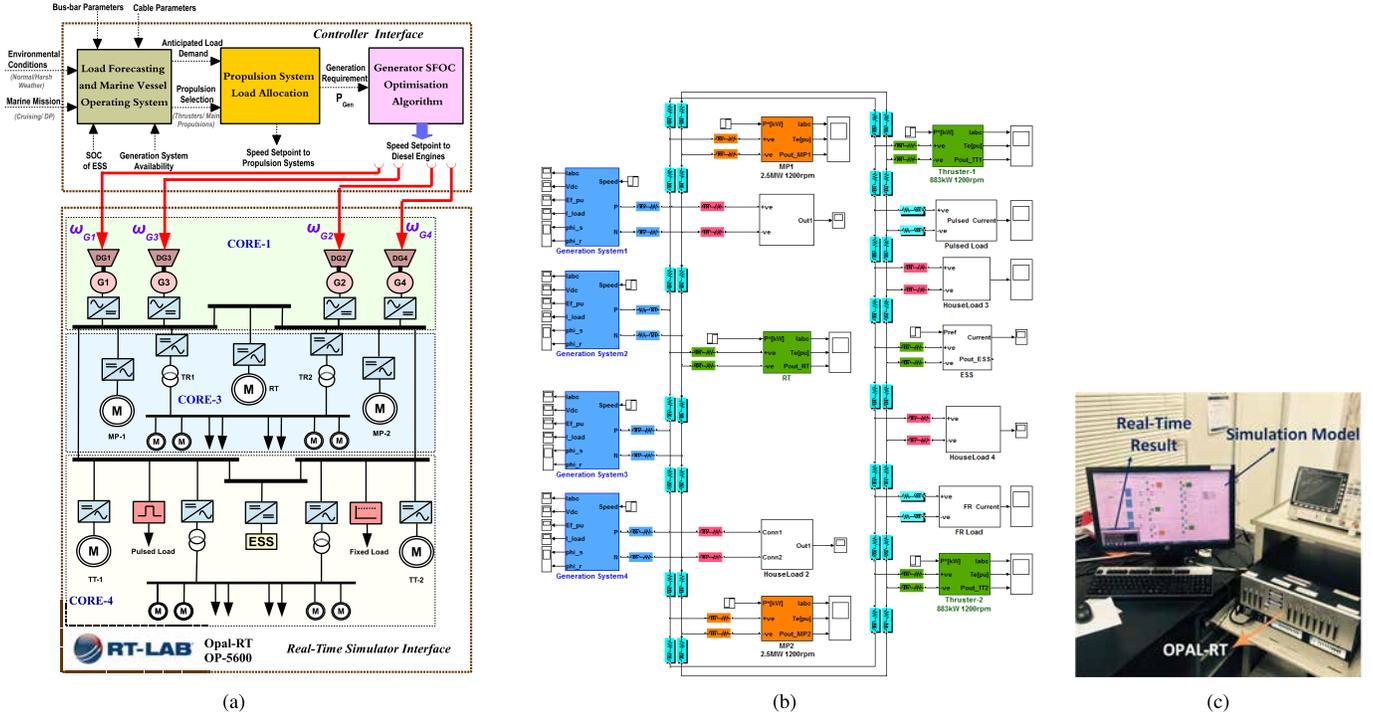

Fig. 9. (a) Architecture of the real-time load scheduling of dc PSV. (b) Schematic in MATLAB/Simulink environment. (c) Representative OPAL-RT setup for real-time simulation.

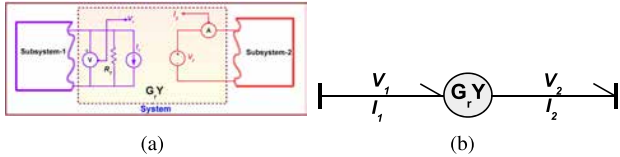

Fig. 10. (a) Representation of gyrators for partioning between subsystem-1 and subsystem-2. (b) Bond graph structure of a gyrator.

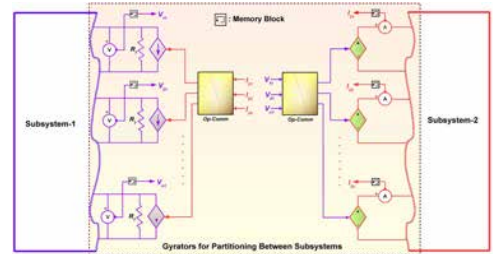

Fig. 11. Gyrator-based system partitioning for "*n*" number of elements.

*1) Overview of Real-Time Simulation System:* The real-time simulation is conducted on the OPAL-RT-based OP5600 real-time simulator which operates on RedHat Linux-based operating system and is interfaced with the host PC by TCP/IP cable. The setup of the real-time simulation system is shown in Fig. 9(c). OPAL-RT uses RT-LAB-based real-time platform, which facilitates the conversion of MATLAB/Simulink models into real-time executable models [55]. It has dedicated toolboxes, such as RT-Events, RTE-drive, and ARTEMiS, to support the real-time simulation system [55], [56]. The execution of the model is achieved by ARTEMiS solver, which is a high-order time-step integration algorithm and is not prone to numerical oscillations [55], [56]. The minimum time step available for real-time simulation in the dc transient real-time simulation model is 10 μs. To comply with such requirements, all the interfaced converters are operated with a switching frequency of 5000 Hz. Furthermore, all the results obtained with the switching models are compared with the averaged models to analyze the performance of VSCs under such time step limitations as well. The partitioning of system using gyrators helps in avoiding numerical inaccuracies by ensuring parallel computation of the partitioned subsystems.

*2) Gyrator-Based Partitioning of System:* Gyrator is an ideal energy transducer used for bond graph representation of a physical system [51]–[53]. This method has been used for partitioning of the bigger marine dc power system into smaller subsystems for parallel computation in real-time transient simulation framework. With reference to Fig. 10(a) and (b), the bigger system is divided into Subsystem-1 and Subsystem-2 with the help of gyrator, $G_rY$ while satisfying the following:

$$V_1 = f(I_2) \tag{34a}$$

$$V_2 = f(I_1). \tag{34b}$$

From 34(a) and 34(b), it can be implied that current $I_2$ in Subsystem-2 is dependent on the voltage $V_1$ of Subsystem-1 or vice versa. This approach can be realized by implementing dependent current and voltage sources. The partitioning of subsystems for "*n*" number of elements utilizing gyrator-based partitioning approach is shown in Fig. 11. In Figs. 10(a) and 11, a very high value resistance ($R_T$) is placed to ensure the numerical consistency of the simulation and memory block is used to avoid algebraic loop errors. The measured current and voltage between the computational subsystems is transferred using `OpComm` block [55], [56]. With the gyrator-based approach, the entire dc marine system is divided into four subsystems (three computational subsystems and one console subsystem). The computational subsystems comprise of one master (`SM_Generator`) and two slave subsystems (`SM_Bus1Load` and `SM_Bus2Load`). The console subsystem (`SC_Console`) is the user interface for data logging. The final partitioned executable file for transient real-time simulation is shown in Fig. 12.

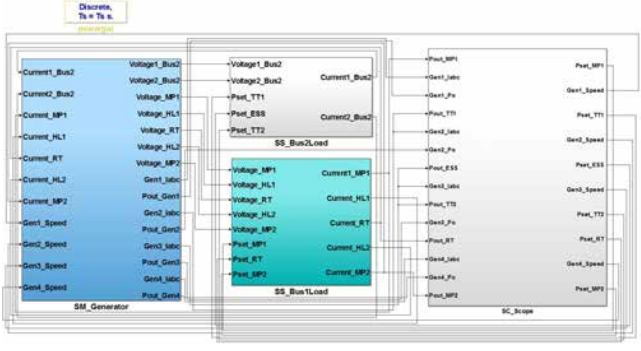

Fig. 12. Representative MATLAB/Simulink model into partitioned subsystems for real-time simulation in OPAL-RT.

*3) Obtaining Results:* The output results from the real-time simulator have been obtained by: 1) monitoring scopes in the console subsystem (`SC_Scope`); 2) by viewing the results in the monitoring oscilloscope; and 3) by saving the data in `.mat` file by `OpWrite` block [55], [56] for offline analysis. All the results presented in this paper are obtained by processing `.mat` files. The results obtained from the oscilloscope are also presented in Section V for comparison with the offline analyzed results.

### D. Algorithm for Real-Time Transient Simulation

Pseudocode for real-time generation scheduling of dc PSV for minimized SFOC is shown in Algorithm 1. It presents the basis for real-time transient simulation scheme in the context of handling various marine missions for minimized SFOC.

## V. Real-Time Simulation Results

With reference to the operational aspects of the proposed dc PSV power management system, this section presents the simulation results of various cases of operating modes and associated contingencies in the real-time operation. The various contingencies associated with dc PSV are listed in Table VI, which has been prepared considering both availability and unavailability of 10% ESS described in Section II-C. From Table VI, it can be observed that the generator and ESS output are marked in red for some specific contingencies. This indicates the overload capabilities of the generation system for supplying high power output for short durations [54]. In the proposed optimization framework, such relaxation has been set for the PSV generation systems to emulate real-time characteristics.

### A. PSV Operating Modes and Associated Contingencies

*1) Dynamic Positioning Operation:* For the DP operation, $\mathbb{L}_{CL}$ is set to zero, while $\mathbb{L}_{HL_{high}}$ and $\mathbb{L}_{HL_{low}}$ are set at 1000 and 270 kVA, respectively, and $\mathbb{L}_{misc}$ is set to the rated value. As described previously, the value of $\mathbb{L}_{DP}$ is dependent on the weather conditions, propeller design, and

---

**Algorithm 1** Pseudocode of Scheduling for DC PSV

1: − Read generation data, network data, load estimation data, static load data, SOC of ESS, and other PSV parameters.
2: − Define operating limits based on the shipboard real-time marine missions. ▷ Eq. (23)-(30).
3: − Build initial Z-bus by handling isolated nodes, if any. ▷ Table II
4: − Create incidence matrix for the existing shipboard network.
5: - Initialize the proposed Optimization Suite. ▷ Eq. (17)-(21)
6: **while** ((Error tolerance for power) < Set limit) **do**
7:    **while** (All options are not processed) **do**
8:      - Find $\triangle X_{bus}$ with power injection matrices
9:      - Calculate power flow, line outage conditions (if any), SOC of ESS
10:      **while** (All scheduling options are not processed) **do**
11:        - Initialize power calculation process of each generator considering different sub-optimal points
12:      **end while**
13:      - Store optimized scheduling options
14:      - Treatment of sub-optimal points. ▷ Section V-C
15:      - Evaluate objective function $f$ ▷ Eq. (6), (13)
16:    **end while**
17:    **for** (Each optimized schedule $P_i$ $i = 1, 2, \ldots$, no. of generators:$P$ **do**
18:      (a) Create speed vector from scheduled generations $P_i$;
19:      (b) Evaluate the schedules for minimized SFOC
20:      Check the error criterion to met. Otherwise, the appropriate speed is chosen from the corresponding generator speed vector.
21:    **end for**
22: **end while**
23: - Print the scheduling plans

---

characteristics. Under normal weather condition, $\mathbb{L}_{DP}$ is set at lower value with all the thrusters operating at lower loading conditions as shown in 35(a). On the contrary at harsh weather conditions, the thrusters are set to be operating at higher loading conditions as shown in 35(b).

$$\mathbb{L}_{DP\ low} = \{\mathscr{L}_{B12}^{TT1}, \mathscr{L}_{B16}^{TT2}, \mathscr{L}_{B5}^{RT}\}$$
$$= \{-300\ \text{kW}, -300\ \text{kW}, -300\ \text{kW}\} \quad (35a)$$
$$\mathbb{L}_{DP\ high} = \{\mathscr{L}_{B12}^{TT1}, \mathscr{L}_{B16}^{TT2}, \mathscr{L}_{B5}^{RT}\}$$
$$= \{-800\ \text{kW}, -800\ \text{kW}, -800\ \text{kW}\}. \quad (35b)$$

Various contingency cases have been prepared, such as loss of generation system and unavailability of ESS (inadequate SOC) at low DP load and high DP load conditions. Utilizing the transient simulation framework formulated in Section IV, the desired power and corresponding operating speed set point of the generation systems $\mathbb{P}_{Gen}$ for all the contingency cases have been listed in Table VI. During the fault at Bus-2 ($\mathscr{P}_{B2}^{Gen24} = 0$) with harsh weather conditions, power demand to the ESS ($\mathscr{P}^{ESS}$) exceeds its rated capacity which cannot be suitably fulfilled. This inadequate generation availability can be mitigated by load shedding operation by

TABLE VI
CONTINGENCY LIST

| CASE | CONTINGENCY CONDITION | CONNECTED LOADS | | | | | Gen-1 | | Gen-2 | | Gen-2 | | Gen-4 | | ESS |
|---|---|---|---|---|---|---|---|---|---|---|---|---|---|---|---|
| | | $TT1$ | $TT2$ | $RT$ | $MP1$ | $MP2$ | $P_{G1}$ | $\omega_{G1}$ | $P_{G2}$ | $\omega_{G2}$ | $P_{G3}$ | $\omega_{G3}$ | $P_{G4}$ | $\omega_{G4}$ | |
| 1A | Sudden gain of DP load | -300 | -300 | -300 | 0 | 0 | +949 | 1130 | +949 | 1130 | +949 | 1130 | +949 | 1130 | +4 |
| | | -800 | -800 | -800 | 0 | 0 | +1324.88 | 1280 | +1324.88 | 1280 | +1324.88 | 1280 | +1324.88 | 1280 | +0.49 |
| 1B | Sudden gain of DP load with inadequate SOC | -300 | -300 | -300 | 0 | 0 | +950 | 1130 | +950 | 1130 | +950 | 1130 | +950 | 1130 | - |
| | | -800 | -800 | -800 | 0 | 0 | +1325 | 1280 | +1325 | 1280 | +1325 | 1280 | +1325 | 1280 | - |
| 2A | Sudden loss of DP load | -800 | -800 | -800 | 0 | 0 | +1324.88 | 1280 | +1324.88 | 1280 | +1324.88 | 1130 | +1324.88 | 1130 | +4 |
| | | -200 | -200 | -200 | 0 | 0 | +899.37 | 1108 | +899.37 | 1108 | +899.37 | 1108 | +899.37 | 1108 | +2.5 |
| 2B | Sudden loss of DP load with inadequate SOC | -800 | -800 | -800 | 0 | 0 | +1325.88 | 1130 | +1325.88 | 1130 | +1325.88 | 1130 | +1325.88 | 1130 | - |
| | | -200 | -200 | -200 | 0 | 0 | +900 | 1109 | +900 | 1109 | +900 | 1109 | +900 | 1109 | - |
| 3A | Bus-2 isolated at low DP load | -300 | -300 | -300 | 0 | 0 | +949 | 1130 | +949 | 1130 | +949 | 1130 | +949 | 1130 | +4 |
| | | -300 | -300 | -300 | 0 | 0 | +1701 | 1425 | +1701 | 1425 | 0 | 0 | 0 | 0 | +398 |
| 3B | Bus-2 isolated at low DP load with inadequate SOC | -300 | -300 | -300 | 0 | 0 | +950 | 1130 | +950 | 1130 | +950 | 1130 | +950 | 1130 | - |
| | | -300 | -300 | -300 | 0 | 0 | +2100 | n/a | +2100 | n/a | 0 | 0 | 0 | 0 | - |
| 4 | Inadequate SOC at low DP load | -300 | -300 | -300 | 0 | 0 | +949 | 1130 | +949 | 1130 | +949 | 1130 | +949 | 1130 | +4 |
| | | -300 | -300 | -300 | 0 | 0 | +950 | 1130 | +950 | 1130 | +950 | 1130 | +950 | 1130 | - |
| 5A | Bus-2 isolated at high DP load | -800 | -800 | -800 | 0 | 0 | +1324.88 | 1280 | +1324.88 | 1280 | +1324.88 | 1280 | +1324.88 | 1280 | +4 |
| | | -800 | -800 | -800 | 0 | 0 | +1934.60 | 1650 | +1934.60 | 1650 | 0 | 0 | 0 | 0 | +1430.8 |
| 5B | Bus-2 isolated at high DP load with inadequate SOC | -800 | -800 | -800 | 0 | 0 | +1325 | 1280 | +1325 | 1280 | +1325 | 1280 | +1325 | 1280 | 0 |
| | | -800 | -800 | -800 | 0 | 0 | +2650 | n/a | +2650 | n/a | 0 | 0 | 0 | 0 | - |
| 6 | Inadequate SOC at high DP load | -800 | -800 | -800 | 0 | 0 | +949 | 1130 | +949 | 1130 | +949 | 1130 | +949 | 1130 | +0.49 |
| | | -800 | -800 | -800 | 0 | 0 | +948.8 | 1130 | +948.8 | 1300 | +948.8 | 1300 | +948.8 | 1300 | - |
| 7A | Sudden Gain in Cruising Load | 0 | 0 | 0 | -1000 | -1000 | +1224.86 | 1243 | +1224.86 | 1243 | +1224.86 | 1243 | +1224.86 | 1243 | +0.55 |
| | | 0 | 0 | 0 | -2500 | -2500 | +1875.17 | 1570 | +1875.14 | 1570 | +1875.14 | 1570 | +1875.14 | 1570 | -399.32 |
| 7B | Sudden Gain in Cruising Load with inadequate SOC | 0 | 0 | 0 | -1000 | -1000 | +1225 | 1243 | +1225 | 1243 | +1225 | 1243 | +1225 | 1243 | - |
| | | 0 | 0 | 0 | -2500 | -2500 | +1975 | 1716 | +1975 | 1716 | +1975 | 1716 | +1975 | 1716 | - |
| 8A | Sudden Loss in Cruising Load | 0 | 0 | 0 | -2500 | -2500 | +1875.17 | 1570 | +1875.17 | 1570 | +1875.17 | 1570 | +1875.17 | 1570 | +399.32 |
| | | 0 | 0 | 0 | -1000 | -1000 | +1224.86 | 1243 | +1224.86 | 1243 | +1224.86 | 1243 | +1224.86 | 1243 | - |
| 8B | Sudden Loss in Cruising Load with inadequate SOC | 0 | 0 | 0 | -2500 | -2500 | +1975 | 1716 | +1975 | 1716 | +1975 | 1716 | +1975 | 1716 | - |
| | | 0 | 0 | 0 | -1000 | -1000 | +1225 | 1243 | +1225 | 1243 | +1225 | 1243 | +1225 | 1243 | - |
| 9 | Bus-2 isolated at low Cruising Load | 0 | 0 | 0 | -1000 | -1000 | +1224.86 | 1243 | +1224.86 | 1243 | +1224.86 | 1243 | +1224.86 | 1243 | 0.55 |
| | | 0 | 0 | 0 | -1000 | -1000 | +1879.62 | 1575 | +1879.62 | 1575 | 0 | 0 | 0 | 0 | +1140.76 |
| 10A | Bus-2 isolated at high Cruising Load | 0 | 0 | 0 | -2500 | -2500 | +1875.17 | 1570 | +1875.17 | 1570 | +1875.17 | 1570 | +1875.17 | 1570 | +399.32 |
| | | 0 | 0 | 0 | -2500 | -2500 | +2750.25 | n/a | +2750.25 | n/a | 0 | 0 | 0 | 0 | +2399.5 |
| 10B | Bus-2 isolated at high Cruising Load (partial load shedding) | 0 | 0 | 0 | -2500 | -2500 | +1875.14 | 1570 | +1875.14 | 1570 | +1875.2 | 1570 | +1875.2 | 1570 | 0.55 |
| | | 0 | 0 | 0 | -2500 | -2500 | +2384.5 | n/a | +2384.5 | n/a | 0 | 0 | 0 | 0 | 3130.96 |
| 10C | Bus-2 isolated at high Cruising Load (maximum load shedding) | 0 | 0 | 0 | -2500 | -2500 | +1875.17 | 1570 | +1875.17 | 1570 | +1875.17 | 1570 | +1875.17 | 1570 | +399.32 |
| | | 0 | 0 | 0 | -2500 | -2500 | +1652.5 | 1400 | +1652.5 | 1400 | 0 | 0 | 0 | 0 | +2014.85 |

TT: tunnel thruster; RT: retractable thruster; MP: main propulsion; Gen: Generator
Output of the generator, ESS and load consumption in kilo watts (kW). Generator shaft speed in *rpm*.

referring to Table I. However, the ESS helps in optimized generation scheduling when the Bus-2 is isolated during low load DP operation and also during sudden gain/loss of DP loads, which is indicated in blue in Table VI. The results pertaining to sudden gain in DP load as per Case 1A of Table VI are shown in Fig. 13. The path of transition of the operating point of the optimized SFOC from initial set point to the final set point has been traced in Fig. 13(d). The trajectory of the SFOC for fixed speed operation is also plotted for comparison with the proposed operation methodology. From Fig. 13(d), it can be inferred that there is substantial reduction of SFOC with the proposed method. The reduction in SFOC is 19% when the DGs are allowed to operate in optimized speed rather than fixed speed during low load DP mode. The operating regime of the DG exceeds the prescribed contour of operation because of the abrupt increase in the load. However, in the real system, the load transition is expected to be smoother rather than sudden abrupt changes. Nevertheless, the initial and final SFOC is optimized and lies within stable region. The output from the monitoring oscilloscope is presented in Fig. 13(e) for comparison with the offline results.

*2) Cruising Mode Operation:* For the cruising operation, $\mathbb{L}_{DP}$ is set to zero, while $\mathbb{L}_{HL_{high}}$, $\mathbb{L}_{HL_{low}}$, and $\mathbb{L}_{misc}$ are set to similar value that of DP operation. The $\mathbb{L}_{CL}$ primarily depends on the operating speed of the vessel. For low cruising speed, the loading of the MP systems is given in 36(a), while for higher speeds, the loading of the MP systems is given in 36(b)

$$\mathbb{L}_{CL\ low} = \{\mathscr{L}_{B3}^{MP1}, \mathscr{L}_{B7}^{MP2}\} = \{-1000\ kW, -1000\ kW\} \tag{36a}$$

$$\mathbb{L}_{CL\ high} = \{\mathscr{L}_{B3}^{MP1}, \mathscr{L}_{B7}^{MP2}\} = \{-2500\ kW, -2500\ kW\}. \tag{36b}$$

Similar to the DP operation, various contingency conditions have been prepared for the cruising loads, which is described in Table VI. In the contingencies associated with fault at Bus-2, $\mathscr{P}_{B14}^{ESS}$ exceeds its rated capacity and thus the implementation of load shedding algorithm becomes impertinent. However, when the Bus-2 is isolated and by setting maximum load shedding by employing $\mathbb{L}_{HL_{high}} == 0$ and $\mathbb{L}_{misc} == 0$, $\mathscr{P}_{B14}^{ESS}$ still exceeds the rated limits as highlighted in Case 10C of Table VI. Thus, during this contingency, higher cruising load cannot be supported by the optimized operation of $\mathbb{P}_{Gen}$, and thus, the speed of the PSV needs to be slowed down to prevent inadvertent black-out condition. However, scheduling of ESS is helpful for the sudden gain/loss of the cruising loads indicated with blue in Table VI. The real-time simulation results during the sudden loss of cruising load are shown in Fig. 14. The transition of SFOC operating point from initial to final set point value is shown in Fig. 14(d). The same operation is repeated when the vessel operates with fixed speed generation system and the SFOC is compared in Fig. 14(d). It can be established that the SFOC is minimized with the proposed method. Although the power is abruptly decreased, the operating point lies within the contours of the operating limits of the DG. As discussed earlier, the change in load would not be abrupt in real scenario, and hence, the DG would operate with optimized SFOC in the stable region. The output from the monitoring oscilloscope is presented in Fig. 14(e) for comparison with the offline results.

### B. Dynamics of the PV-Based BESS System

*1) Charging of BESS:* As discussed in Section II-C, the charging of the BESS could be carried out either by slow charging or fast charging schemes. Fig. 15(a) shows

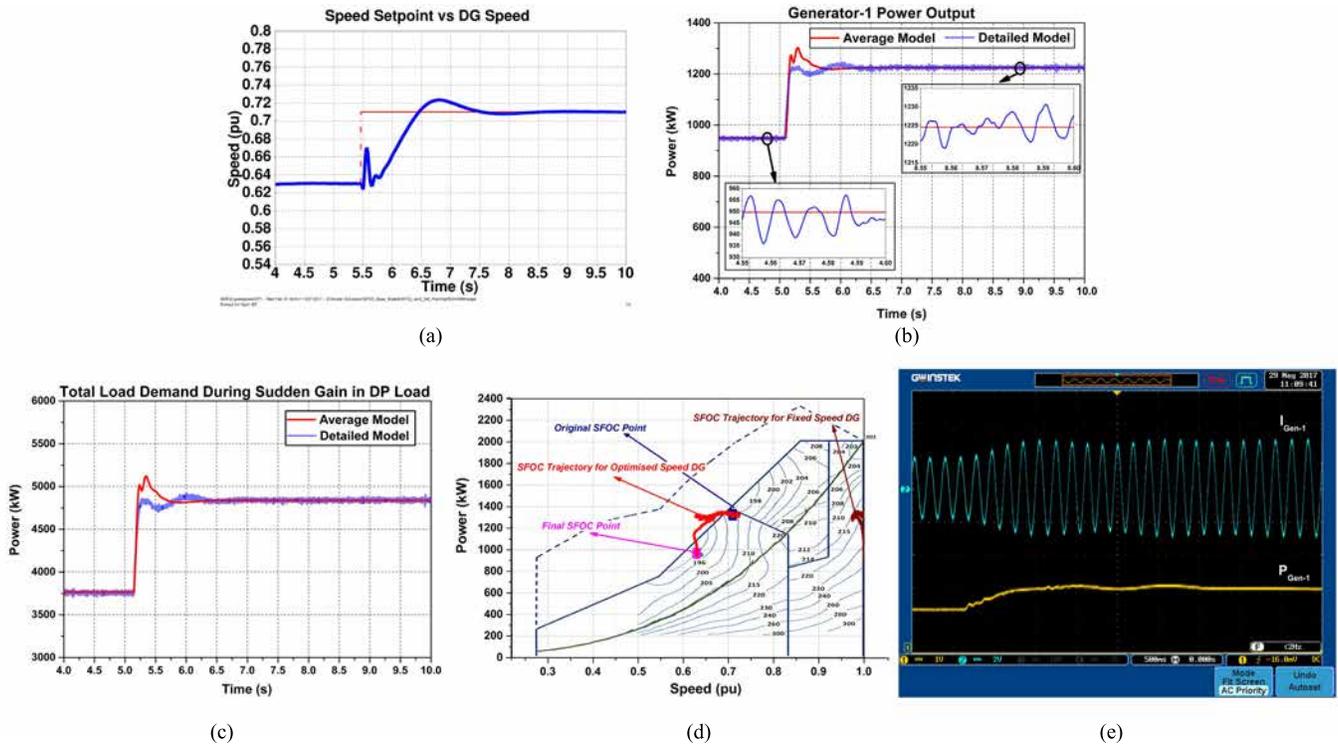

Fig. 13. (a) Speed set point versus DG speed. (b) Power variation of DG-1. (c) Change in total load demand. (d) Transition of SFOC to optimized point in real-time. (e) Output in monitoring oscilloscope during sudden gain in DP load as per Case 1A of Table VI.

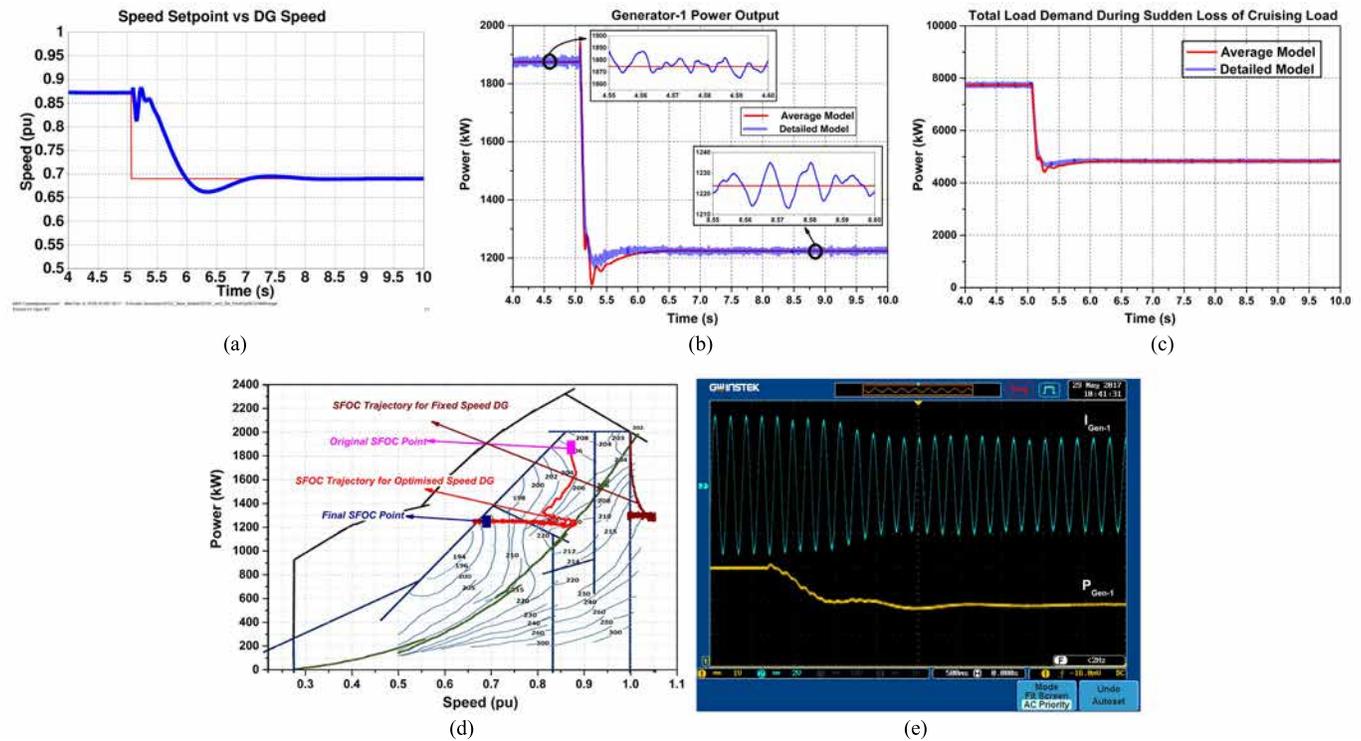

Fig. 14. (a) Speed set point versus DG speed. (b) Power variation of DG-1. (c) Change in total load demand. (d) Transition of SFOC to optimized point in real time. (e) Output in monitoring oscilloscope during sudden loss in cruising load as per Case 8A of Table VI.

the variation of irradiance and the output power of the dc/dc-1 while operating at MPPT mode of operation. The PV current ($i_{pv}$) for charging the BESS is shown in Fig. 15(b). For fast charging, the dc/dc-2 maintains battery charging current determined by the set point ($i_{batt-SP}$) as shown in Fig. 15(c). With the

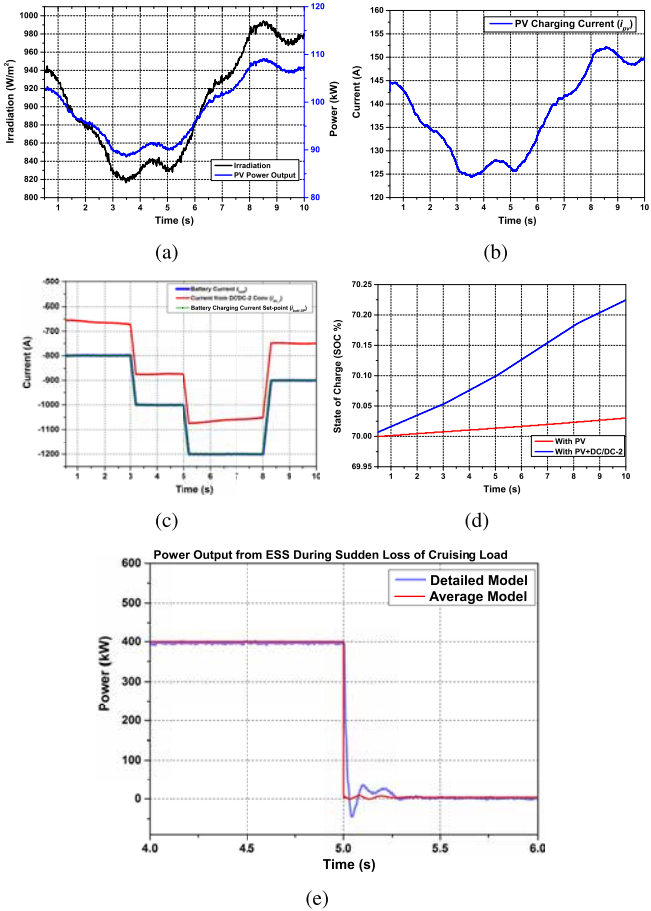

Fig. 15. (a) Change in power availability from PV panels with change in irradiation. (b) Variation of current output of PV panel while operating it in MPPT mode. (c) Charging current characteristic from dc/dc-2. (d) Change in SOC of BESS when it is charged from PV panel and both with PV panel and dc/dc-2. (e) DC/DC-2 output for contingency 8A in Table VI.

dc/dc-1 current $i_{pv}$, the variation of $i_{dc\_i}$ and $i_{batt}$ with change of irradiation and battery charging current set point is shown in Fig. 15(c). Fig. 15(d) shows the variation of SOC of the BESS when it is subjected to slow charging and fast charging, respectively. The slow charging of the BESS can be employed, while the PSV is in dockyard. Alternatively, the fast charging schemes might be employed by forecasting the nature of job to be done by the PSV and if the job requires intermitted power requirements.

2) *Discharging of BESS:* The discharging of the BESS is explained with reference to Case 8A of Table VI. During sudden change of the DP operation, the power demand of the BESS decreases from 399.32 to 0.55 kW. The power delivered by the PV-BESS system is shown in Fig. 15(e).

### C. Treatment of Suboptimal Points

Treatment of suboptimal points has been explained with respect to the contingency scenario 8A of Table VI, where the initial derived generation schedule of 1875.17 kW and ESS of 399.32 kW are the points-1 in Fig. 16(a) and Table VII at which the SFOC is calculated. Such an SFOC is dependent on the operating speed of the DG and the ESS output. However, it has been observed that the scheduling constraints are also satisfied at the suboptimal locations with certain generator schedules and corresponding SFOC values associated with it. These set of points are termed as *local optimum* points. To illustrate this, $\mathscr{P}_G^n$ is considered as generator schedule and $f$ as the objective function used in Algorithm 1. Equations 37(a) and 37(b) represent the conditions for *minima* [26]

$$\mathscr{P}_G^n \in \mathbb{R} \implies f'(P_G^n) = 0 \tag{37a}$$
$$f'(\mathscr{P}_G^n) > 0. \tag{37b}$$

Conditions for local minima (suboptimal point) is given in the following:

$$f^* = f(\mathscr{P}_G^{n*}) \tag{38}$$

for local minimizer $\mathscr{P}_G^{n*}$. This is the smallest function value in some feasible neighborhood defined by the following:

$$\mathscr{P}_G^{n*} \in \Omega. \tag{39}$$

There exists a $\delta > 0$ such that

$$f^* \leqslant f(\mathscr{P}_G^{n*}) \quad \forall \ P_G^n \tag{40a}$$
$$\text{in } \left\{ \mathscr{P}_G^{n*} \in \Omega : |\mathscr{P}_G^n - \mathscr{P}_G^{n*}| \right\} \leqslant \delta. \tag{40b}$$

Thus, there can be many local minima, i.e., multiple suboptimal points, which are not global minima. Special properties such as the convexity of feasible region "$\Omega$" and objective function "$f$" imply that any local solution is a global solution. It has been observed that during real-time simulations, such suboptimal (local optimum points), which satisfy the given constraints, has the potential to speed up the calculations of the scheduling process while diligently treating the associated ESS to arrive at better SFOC values than the corresponding SFOC at global optima. This has been demonstrated using Table VII corresponding to Fig. 16(a), where the ESS is varied at all suboptimal locations to study the impact on the SFOC. The absence of ESS during the initial state at $P_{G1}$ would improve the SFOC operating scenario (*Point-2*), however, the limits of generator capacity (2048 kW) are enforced to consider the ESS as an option (*Point-1*). Here, ESS is delivering at 399.32 kW and the corresponding SFOC is 207 (*Point-1*). Now, with the sudden loss of the cruising load, we could notice that before arriving to the global optimum point $P_{G2(g)}$, it has been passing through a local optima $P_{G2(l)}$ (*Point-3*) where the corresponding SFOC is 205 against the global optimum SFOC of 197 ($P_{G2(g)}$, *Point-5*). During such a phase, it is evident that ESS is no longer required to contribute to the load, and accordingly, scheduling algorithm suggested the output of ESS at +0.55 kW. However, considering the enforced operation of ESS, at this instant, shall make the corresponding SFOCs of $P_{G2}$ and $P_{G2(l)}$ at 195 (*Point-6*) and 200 (*Point-4*), respectively. Hence, instead of abrupt suspension of ESS supply, slowly adjusting the ESS in such a way to yield the best scenario of SFOC is a possibility and where the suboptimal points can be diligently treated.

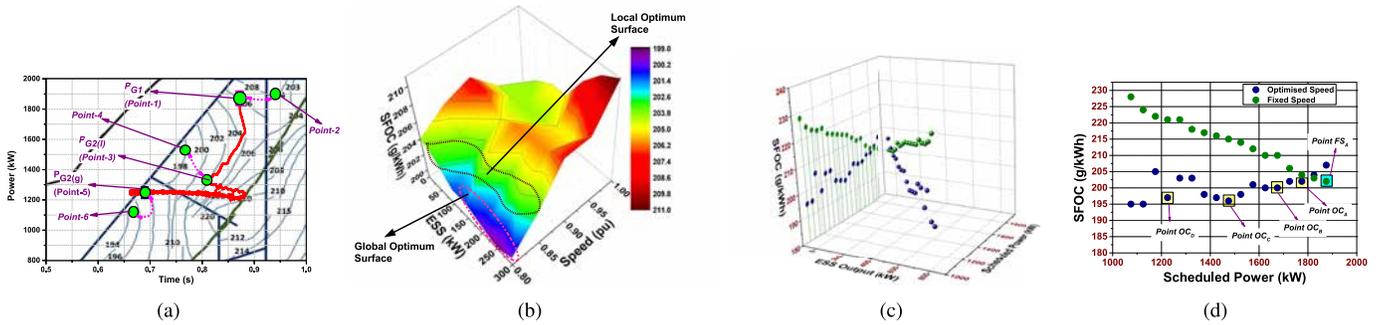

Fig. 16. (a) Trajectory of the SFOC during the sudden loss of cruising load with locations of various suboptimal points. (b) Surface plot of SFOC variation for varying ESS and DG speed. (c) Comparison of the changing operating points. (d) Location of suboptimal points for DGs operating at optimized speed and fixed speed.

TABLE VII
VARIATION OF SFOC AT SUBOPTIMAL POINTS DURING
THE SUDDEN LOSS OF CRUISING LOAD

| Points | ESS Output (kW) | Generation Output (kW) | Optimised Speed (p.u.) | SFOC (g/kWh) |
|--------|-----------------|------------------------|------------------------|--------------|
| 1 | 400 | 1875 | 0.872 | 207 |
| 2 | 0 | 1975 | 0.953 | 204 |
| 3 | 0 | 1330 | 0.72 | 205 |
| 4 | 400 | 1230 | 0.70 | 200 |
| 5 | 0.55 | 1225 | 0.69 | 197 |
| 6 | 400 | 1125 | 0.66 | 195 |

Table VII shows the variation of SFOC corresponding to these suboptimal points considering ESS in operation even after the sudden loss of load and at optimized speed of generators. However, it may not be prudent to consider such points in all the contingency cases owing to operational constraints such as availability and SOC of the interfaced ESS. So, it is evident that the suboptimal point can yield better SFOC at optimized speed and further it also has an upper hand in reducing the time propagation of schedules for feeding to the shipboard controllers.

So, in the proposed work, it was noticed that suboptimal points can accommodate ESS for better performance of DGs. To speed up the calculation in scheduling process and to treat the consideration of ESS, suboptimal schedules can yield a feasible solution.

Dependence of ESS output on suboptimal locations is further demonstrated with the help of Fig. 16(b)–(d). In Fig. 16(b), power extraction from ESS is varied from 0 to 300 kW to visualize the changing operating points of the SFOC of DG while fulfilling constant load demand of 1875 kW. The location of the global optima with minimized SFOC and local optima is highlighted in Fig. 16(b). In Fig. 16(c) and (d), it is further compared with the DG operating at constant speed. With reference to Fig. 16, the following observations can be cited as follows.

1) Fig. 16(b) shows the surface plot of the SFOC variation with ESS and DG speed in the aforesaid treatment of suboptimal points. The suboptimal scheduled points are also marked in the figure.
2) Fig. 16(c) and (d) shows the comparative analysis of the location of suboptimal points for fixed speed and optimized speed operation of the DGs. It can be observed that the minimum SFOC for the corresponding suboptimal points for fixed speed DGs is achieved while operating at rated conditions denoted by *Point $FS_A$* in Fig. 16(d).
3) For the DGs operating at optimized speed, there are multiple suboptimal points depending on ESS output and operating speed of DG. These suboptimal points, the variable speed DGs, and the suboptimal locations are dependent on the injected power from ESS and DG operating speed as shown in Fig. 16(d). The location of the suboptimal points is shown as *Point $OS_A$*, *Point $OS_B$*, *Point $OS_C$*, and *Point $OS_D$* in Fig. 16(d).

### D. Comparison With the Operation at Different Speeds

For all the operating cases in Table VI, SFOC is compared when the generator is running at 1800 rpm and 1600 rpm against the calculated optimized speed. The variation of the operating points is shown in Fig. 17. It can be seen that SFOC is significantly reduced when run at optimized speed, thus highlighting the advantages of the proposed real-time transient simulation scheme and corresponding optimization framework.

### E. Influence of ESS on SFOC

From Section V-C, it is evident that ESS has strategic influence on various marine missions not only on dealing with short time-load transients for smooth operations but also on improving the transient responses of DGs and corresponding SFOC. The set point to ESS and DG output have been decided based on the cost functions, cable loadings, and the realistic criteria discussed in Section IV. From Fig. 17(b), it is evident

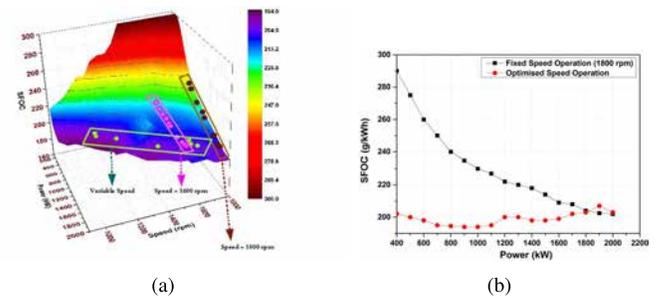

Fig. 17. (a) Variation of the operating point of SFOC for generator running at 1800 rpm, 1600 rpm, and variable speed (optimized speed). (b) Comparison of SFOC for generator running at 1800 rpm and optimized speed.

TABLE VIII
INFLUENCE OF ESS ON SFOC DURING THE SUDDEN LOSS OF CRUISING LOAD

| WITHOUT ESS | | | | | | | | | WITH ESS | | | | | | | | |
|---|---|---|---|---|---|---|---|---|---|---|---|---|---|---|---|---|---|
| $P_{G1}$ | $\omega_{G1}$ | $P_{G2}$ | $\omega_{G2}$ | $P_{G3}$ | $\omega_{G3}$ | $P_{G4}$ | $\omega_{G4}$ | SFOC (CS/OS) | $P_{G1}$ | $\omega_{G1}$ | $P_{G2}$ | $\omega_{G2}$ | $P_{G3}$ | $\omega_{G3}$ | $P_{G4}$ | $\omega_{G4}$ | SFOC (CS/OS) |
| +1974.97 | 1716 | +1974.97 | 1716 | +1974.97 | 1716 | +1974.97 | 1716 | 203/204 | +1875.17 | 1570 | +1875.17 | 1570 | +1875.17 | 1570 | +1875.17 | 1570 | 203/204 |
| +1224.99 | 1243 | +1224.99 | 1243 | +1224.99 | 1243 | +1224.99 | 1243 | 222/200 | +1224.86 | 1243 | +1224.86 | 1243 | +1224.86 | 1243 | +1224.86 | 1243 | 222/200 |

CS: Constant Speed (1800 rpm); OS: Optimised Speed

that at higher loads, the SFOC is almost constant for optimized variable speed and constant speed operation with or without ESS and the same has been shown in Table VIII. However, having ESS during generators meeting higher loads can reduce the initial speed deviations and torsional stress and further it enables the option of catering any unexpected load demand.

## VI. CONCLUSION

This paper investigated the modeling and control approaches of dc PSV with real-time transient simulation framework to minimize the fuel consumption in terms of SFOC while considering the treatment of suboptimal points along with an option of utilizing onboard PV-based energy storage facility. This sort of real-time operation is expected to be a key ingredient of the future autonomous marine vehicles.

1) DC OPF-based algorithms have been applied for real-time scheduling of generation resources with an objective to minimize the fuel consumption. This operation is within the framework of the proposed real-time transient simulation setup and has been successfully demonstrated in this paper. Results obtained are promising and strengthens the ability of the future dc marine vessels to comply with the upcoming stringent laws on pollution control. It has been shown that the output of ESS can influence the generator set points particularly during sudden load changes for DP and cruising missions.

2) Traditionally, the power sharing happens according to generator ratings through droop control rather than optimal generation scheduling as per load demand for a particular marine mission. Hence, an optimal scheduling system based on dc OPF for tackling any specific nature of marine missions has been demonstrated in this paper and the results indicated such an approach will enable the PSV to operate in fuel efficient regime with improved transient responses.

3) Responses of the DGs to various contingencies are studied and it has been found that unavailability of the generation system due to fault at bus bar demands higher requirement of ESS to satisfy the load demands. In such cases, it is pertinent to have load shedding routines, which are dependent on the marine missions to prevent power blackout of the vessel. This highlights the necessity for considering such routines in future autonomous dc PSVs. Some cases have been highlighted using the proposed real-time transient simulation framework.

4) Higher fuel savings are noticed while the generators are operated at optimized speed for low power demand. In comparison with fixed speed operation, reduction in SFOC of 19% has been reported when the PSV was operating at low-power DP operation.

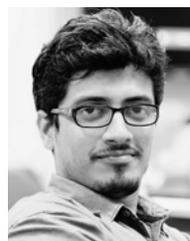

**Kuntal Satpathi** (S'14) received the B.Tech. degree in electrical engineering from the Haldia Institute of Technology, Haldia, India, in 2011. He is currently pursuing the Ph.D. degree with the School of Electrical and Electronic Engineering, Nanyang Technological University, Singapore.

From 2011 to 2014, he was with Jindal Power Ltd., Raigarh, India, specializing in power plant operations. His current research interest includes modeling, control and protection of dc grids, and power electronics for dc distribution system.

778

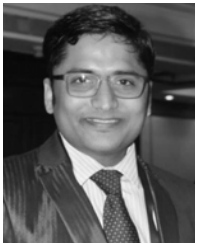

**VSK Murthy Balijepalli** (M'06) received the Ph.D. degree from IIT Bombay, Mumbai, India, in 2014.

He was with India Smart Grid Task Force, Ministry of Power, New Delhi, India, the Inter-Ministerial Group on Smart Grids (The Future of Energy), from 2014 to 2016, where he was involved in the project execution and evaluation of smart grid pilot projects in India. Since 2016, he has been a Research Fellow with School of Electrical and Electronic Engineering, Nanyang Technological University, Singapore. His current research interests include power system modeling and optimization, microgrid resiliency, policy making, and model building for emerging power systems and smart grids.

Dr. Balijepalli was an Active Member of various technical panels under the Bureau of Indian Standards (LITD-10), CIMug, and IEC PC118-Smart Grid User Interface. He is the Founder of DesiSmartGrid.com, India's first Smart Grid Educational Portal. He was a recipient of the Massachusetts Institute of Technology Young Innovator Award, the Department of Science and Technology-Lockheed Martin Gold Medal, the Institute of Engineers India Young Engineer Award (U.K. Royal Charter of Incorporation), and the Gandhian Technological Edge Award for his outstanding research on smart grids. He is currently serving on the expert committee of Sustainable Energy to the United Nations Economic Commission for Europe, Geneva, and Switzerland.

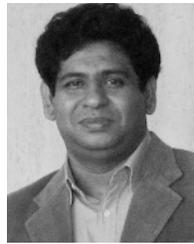

**Abhisek Ukil** (S'05–M'06–SM'10) received the B.E. degree in electrical engineering from Jadavpur University, Kolkata, India, in 2000, the M.Sc. degree in electronic systems and engineering management from the University of Bolton, Bolton, U.K., in 2004, and the Ph.D. degree, with a focus on automated disturbance analysis in power systems, from the Tshwane University of Technology, Pretoria, South Africa, in 2006.

From 2006 to 2013, he was a Principal Scientist with the ABB Corporate Research Center, Baden, Switzerland, where he led several projects on smart grid, protection, control, condition monitoring, including the first worldwide prototype of directional protection relay using only current for smart grid applications. Since 2013, he has been an Assistant Professor with the School of Electrical and Electronic Engineering, Nanyang Technological University, Singapore, where he has been leading a group of 20 researchers with several industrial collaborations. He has authored over 125 refereed papers, a monograph, and two chapters. He has invented ten patents. His current research interests include smart grid, dc grid, protection and control, renewable energy and integration, energy storage, and condition monitoring.